\documentclass[amsmath,amssymb,superscriptaddress,nobalancelastpage,prb,preprint]{revtex4-1}
\usepackage{hyphenat}
\usepackage{graphicx,xcolor}
\usepackage{graphicx}
\usepackage{varioref}
\usepackage{xr-hyper}
\usepackage{xcolor}
\usepackage{nicefrac}
\usepackage{xfrac}
\usepackage{hyperref}
\hypersetup{colorlinks,linkcolor=blue,urlcolor=blue,citecolor=blue}
\usepackage{ulem}
\usepackage{siunitx}
\usepackage{graphicx}
\usepackage{dcolumn}
\usepackage{bm}
\usepackage{braket}
\usepackage{wasysym}
\usepackage{textcomp}
\usepackage{mhchem}

\newcommand{\DSF}[2]{{\ensuremath{\displaystyle{\frac{#1}{#2}}}}}

\newcommand{\Sinc}{{\ensuremath{\mathrm{sinc}}}}
\newcommand{\sinc}[1]{{\ensuremath{\mathrm{sinc}\left({#1}\right)}}}
\newcommand{\IMA}{{\ensuremath{\mathrm{i}}}}
\newcommand{\DD}[1]{{\ensuremath{\mathrm{d}{#1}}}}
\newcommand{\DDD}[3]{{\ensuremath{\mathrm{d}^{#1}{#2}}}}

\newcommand{\eref}[1]{Eq.~(\ref{#1})}

\newcommand{\eeref}[2]{Eqs.~(\ref{#1},\ref{#2})}

\newcommand{\sref}[1]{Sec.~\ref{#1}}
\newcommand{\cref}[1]{Chap.~\ref{#1}}

\newcommand{\fref}[1]{Fig.~\ref{#1}}

\newcommand{\vq}{\ensuremath{\bm{q}}}

\newcommand{\vkr}{\ensuremath{\bm{r}}}
\newcommand{\vk}{\ensuremath{\bm{k}}}

\newcommand{\lrb}[1]{\ensuremath{\left({#1}\right)}}
\newcommand{\lrs}[1]{\ensuremath{\left[{#1}\right]}}

\newcommand{\lrv}[1]{\ensuremath{\left|{#1}\right|}}

\newcommand{\lra}[1]{\ensuremath{\left\langle{#1}\right\rangle}}
\newcommand{\EE}{\ensuremath{\mathrm{e}}}

\begin{document}



\title{Diffraction from nanocrystal superlattices}

\author{Antonio Cervellino}
\email{antonio.cervellino@psi.ch}
\affiliation{Swiss Light Source, Paul Scherrer Institut, Villigen, Switzerland}

\author{Ruggero Frison}
\affiliation{Physik-Institut, Universit\"{a}t Z\"{u}rich, Winterthurerstrasse 190, CH-8057 Z\"{u}rich, Switzerland}


\begin{abstract}
Diffraction from a lattice of periodically spaced crystals is a topic of current interest because of the great development of self-organised superlattices (SL) of nanocrystals (NC). The self-organisation of NC into SL has theoretical interest, but especially a rich application prospect, as the coherent organisation has large effects on a wide range of material properties. Diffraction is a key method to understand the type and quality of SL ordering. Hereby the characteristic diffraction signature of a SL of NC - together with the characteristic types of disorder - are theoretically explored.
\end{abstract}


\maketitle



\section{Introduction}

We will explore the diffraction characteristics of supercrystals (SCs) as superlattices (SLs) of nanocrystals (NCs) where 
the periodic entity is itself a small crystal. There is a widespread current interest on such material, 
driven by 
the changes in properties that the periodic organisation of NCs into a SC yields. 
The synthesis of SCs is a result of the always more sophisticated ways of synthesising NCs (and nanoparticles in general) with very sharp distributions in size and well defined faceted shape. These NCs then, under proper conditions, self-organise forming a SL and thence a SC. 
A brief current perspective on SCs and their properties can be found in \cite{Santos_2021}, with 
diffraction signatures discussed in \cite{Bertolotti_2022} and references therein. Here we would 
deepen the discussion of diffraction theory of SCs. In particular, we will focus on the diffraction signature 
of imperfectly ordered SCs, especially concerning the effect of NCs of different sizes in the SL nodes, 
and the effect of slight rotation of the component NCs with respect to each other. The latter part will be developed only partly in this paper, delegating the full discussion to an upcoming technical paper. 

\section{Materials and Methods}

Some of the numerical simulations hereby presented were computed using the 
DEBUSSY software suite \cite{Cervellino:to5122} and \emph{ad hoc} written code in Fortran2008 (available by email from the authors) and the ZODS program \cite{frison_zods_2016}.

\section{Perfect superlattice of identical nanocrystals}\label{sec:perfectSC}

A perfect superlattice of identical nanocrystals, perfectly equioriented in space and periodically arranged without any defect, is clearly a nonissue - it can be dealt with as a conventional crystalline structure with a large unit cell. However, there is also an interesting and simple analytic formula describing the diffraction amplitude 
of such supercrystal, if some inessential shape restrictions are assumed. 
We will consider parallelohedral nanocrystals, extended along the unit cell vectors $\vk{a}$,  $\vk{b}$, 
 $\vk{c}$, and whose nanocrystal lattice coordinates are defined by integers $n_a,n_b,n_c$:
 \begin{equation}
 \lrb{n_a\vk{a},n_b\vk{b},n_c\vk{c}} \ \Bigl|\Bigr.\ 0\leqslant n_a \leqslant N_a, \ 0\leqslant n_b \leqslant N_b,\ 0\leqslant n_c \leqslant N_c.
 \label{eq:parall1}
 \end{equation}
 The superlattice cell vectors we suppose to be direct multiples of the crystal cell vectors:
 \begin{eqnarray}
\vk{a}_S&=&(N_a+s_a)\vk{a};\nonumber\\
\vk{b}_S&=&(N_b+s_b)\vk{b};\label{eq:spacy}\\
\vk{c}_S&=&(N_c+s_c)\vk{c}.\nonumber
\end{eqnarray}
The spacings constants $s_a,s_b,s_c$ are supposed to be positive, otherwise we'd have coalescence (or even overlap) of the crystal domains. Coalescence would bring us to polycrystalline matter, that is quite another issue. Instead, the spacings are supposed to be filled by some kind of ligand. 
Again, we assume a parallelohedral shape for the supercrystals, assuming that the occupied superlattice nodes 
lie at integer multiples of $\vk{a}_S,\vk{b}_S,\vk{c}_S$, similarly to Eq.~\ref{eq:parall1}:
\begin{equation}
 \lrb{m_a\vk{a}_S,m_b\vk{b}_S,m_c\vk{c}_S} \ \Bigl|\Bigr.\ 0\leqslant m_a \leqslant M_a, \ 0\leqslant m_b \leqslant M_b,\ 0\leqslant m_c \leqslant M_c.
 \label{eq:parallS}
 \end{equation}

We abstain in the following from describing the atomic content of the unit cell; we will assume 
that each NC unit cell contains just one point scatterer of unit scattering power in the origin. 
Generalisation to real NCs with a specified unit cell content is straightforward but able to unnecessarily complicate the notation. The NC's scattering density is described formally in Eq.~\ref{eq:rhoNC}.

\begin{figure}[!tbh]
\centering
 \includegraphics[width=10.5cm]{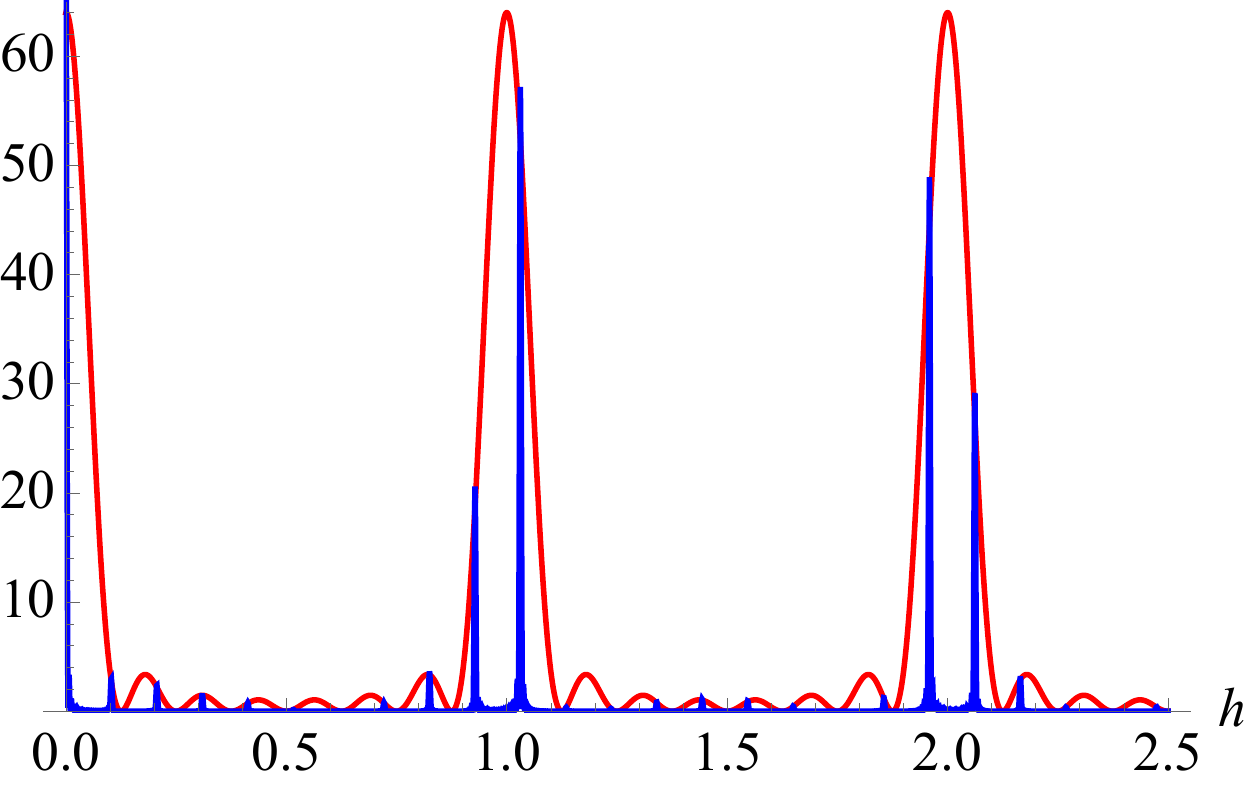}
\caption{Red line: the  NC scattered intensity along $\vq=h\vk{a}^*$, $F_{NC}^2(h\vk{a}^*)$, for $N_a=8$. Note a sharp peak at all integer values of $h$. 
The peaks have all the same shape and are bracketed by zeroes at $h=k\pm 1/N_a$, $k\in\mathbb{Z}$. 
Factors $M_b,M_c$ 
Blue line: $I_{SC}(\vq)/M_a^2$ for $N_a=8$, $M_a=20$, $\vk{a}_S=9.7\vk{a}$. One can clearly see the very sharp SL peaks modulated by the NC scattered intensity, so that each NC peak is replaced by a tight "copse" of sharper SL peaks.}
\label{fig:Laue}
\end{figure}

The unit cell of the superlattice contains instead a single NC. We can arbitrarily set each superlattice node in the NC's scattering barycentrum 
\[
\vk{C}=\lrb{(N_a+1)\vk{a},(N_b+1)\vk{b},(N_c+1)\vk{c}}/2.
\] 
As such, the scattering density of a SC is just that of the SL (with - again - unit power point scatterers on the lattice nodes, see Eq.~\ref{eq:rhoSL}) {\emph{convoluted}} with the scattering density of a NC (Eq.~\ref{eq:rhoSC}). 
\begin{eqnarray}
\rho_{NC}(\vkr) &=& 
\mathop{\sum}_{n_a=1}^{N_a}\mathop{\sum}_{n_b=1}^{N_b}\mathop{\sum}_{n_c=1}^{N_c}
\delta\lrb{\vkr- \lrb{n_a\vk{a},n_b\vk{b},n_c\vk{c}} +\vk{C}};\label{eq:rhoNC}\\
\rho_{SL}(\vkr) &=& 
\mathop{\sum}_{m_a=1}^{M_a}\mathop{\sum}_{m_b=1}^{M_b}\mathop{\sum}_{m_c=1}^{M_c}
\delta\lrb{\vkr- \lrb{m_a\vk{a}_S,m_b\vk{b}_S,m_c\vk{c}_S} };\label{eq:rhoSL}\\
\rho_{SC}(\vkr) &=& \int_{\mathbb{R}^3}\DDD{3}{\vkr'}\rho_{NC}(\vkr')\rho_{SL}(\vkr-\vkr')\nonumber\\
&=&\!\!\!
\mathop{\sum}_{m_a=1}^{M_a}\mathop{\sum}_{m_b=1}^{M_b}\mathop{\sum}_{m_c=1}^{M_c}
\mathop{\sum}_{n_a=1}^{N_a}\mathop{\sum}_{n_b=1}^{N_b}\mathop{\sum}_{n_c=1}^{N_c}
\!\!\delta\lrb{\vkr-\! \lrb{n_a\vk{a},n_b\vk{b},n_c\vk{c}} \!+\!\vk{C}\!-\!\lrb{m_a\vk{a}_S,m_b\vk{b}_S,m_c\vk{c}_S} 
\!}\label{eq:rhoSC}
\end{eqnarray}
It follows that the SC's scattering amplitude (the Fourier transform) is the product of the scattering amplitude of a NC times that of a SL decorated with unit point scatterers. 
\begin{eqnarray}
F_{NC}(\vq) &=& 
\int_{\mathbb{R}^3}\DDD{3}{\vkr}\rho_{NC}(\vkr)\EE^{2\pi\IMA \vq\cdot\vkr}
\,;\label{eq:FNC}\\
F_{SL}(\vq) &=& 
\int_{\mathbb{R}^3}\DDD{3}{\vkr}\rho_{SL}(\vkr)\EE^{2\pi\IMA \vq\cdot\vkr}\,;\label{eq:FSL}\\
F_{SC}(\vq) &=& \int_{\mathbb{R}^3}\DDD{3}{\vkr}\rho_{SC}(\vkr)\EE^{2\pi\IMA \vq\cdot\vkr}
=F_{NC}(\vq)F_{SL}(\vq)
\label{eq:FSC}
\end{eqnarray}
Here $\vq$ is the transferred momentum vector, whose length is $q=\lrv{\vq}=2\sin(\theta)/\lambda$, with $\lambda$ the incident wavelength and $\theta$  half of the deflection angle. 
The transform in Eq.~{eq:FNC} has been historically evaluated by Max von Laue \cite{FriedrichKnippingLaue1912,Laue1912}, as
\begin{equation}\label{eq:LaueANC}
F_{NC}(\vq)=\DSF{\sin\lrb{N_a\pi\vq\cdot\vk{a}}}{\sin\lrb{\pi\vq\cdot\vk{a}}}
\DSF{\sin\lrb{N_b\pi\vq\cdot\vk{b}}}{\sin\lrb{\pi\vq\cdot\vk{b}}}
\DSF{\sin\lrb{N_c\pi\vq\cdot\vk{c}}}{\sin\lrb{\pi\vq\cdot\vk{c}}}
\end{equation}
In more modern form, using the Chebyshev polynomials of the second kind $U_k(x)$ (see \cite{Wolfram-Ucheb}, Eq.~(22)), we can rewrite it as
\begin{equation}
F_{NC}(\vq)=
U_{N_a-1}\lrb{\cos\lrb{\pi\vq\cdot\vk{a}}}
U_{N_b-1}\lrb{\cos\lrb{\pi\vq\cdot\vk{b}}}
U_{N_c-1}\lrb{\cos\lrb{\pi\vq\cdot\vk{c}}}
\end{equation}
Similarly, 
\begin{equation}
F_{SL}(\vq)=\EE^{-2\pi \IMA \vq\cdot\vk{C}_S}
U_{M_a-1}\lrb{\cos\lrb{\pi\vq\cdot\vk{a_S}}}
U_{M_b-1}\lrb{\cos\lrb{\pi\vq\cdot\vk{b_S}}}
U_{M_c-1}\lrb{\cos\lrb{\pi\vq\cdot\vk{c_S}}}
\end{equation}
The phase factor is because we have not referred our SL slab to its scattering barycentrum 
\[
\vk{C}_S=\lrb{(M_a+1)\vk{a}_S,(M_b+1)\vk{b}_S,(M_c+1)\vk{c}_S}/2
\]
but it is inessential. 
In fact, to obtain the scattered intensity $I_{SC}(\vq)$, we take the square modulus of $F_{SC}(\vq)$, 
\begin{equation}
I_{SC}(\vq) =\lrv{ F_{NC}(\vq)F_{SL}(\vq) }^2=F_{NC}^2(\vq)\lrv{F_{SL}(\vq) }^2
\end{equation}
where the phase factor disappears and we have a product of six squared Chebyshev polynomials. 
A simple graph shows these simple functions for $\vq=h\vk{a}^*$, along the NC reciprocal axis $\vk{a}^*$. 
The reciprocal space vectors are defined by 
\[
\vk{a}^*\cdot\vk{a}=\vk{b}^*\cdot\vk{b}=\vk{c}^*\cdot\vk{c}=1;\qquad
\vk{a}^*\cdot\vk{b}=\vk{b}^*\cdot\vk{c}=\vk{c}^*\cdot\vk{a}=\vk{a}^*\cdot\vk{c}=\vk{b}^*\cdot\vk{a}=\vk{c}^*\cdot\vk{b}=0.
\]
We also assume - for this example - that the SL vectors 
\[
\vk{a}_S\propto \vk{a};\qquad\vk{b}_S\propto \vk{b};\qquad\vk{c}_S\propto \vk{c}.
\]
Therefore if $\vq=h\vk{a}^*$ then 
\[
\cos\lrb{\pi\vq\cdot\vk{b}}=\cos\lrb{\pi\vq\cdot\vk{c}}
\cos\lrb{\pi\vq\cdot\vk{b_S}}=\cos\lrb{\pi\vq\cdot\vk{c_S}}=\cos(0)=1
\]
and 
\[
U_{N_b-1}\lrb{1}=N_b;\qquad U_{N_c-1}\lrb{1}=N_c;\qquad
U_{M_b-1}\lrb{1}=M_b;\qquad U_{M_c-1}\lrb{1}=M_c.
\]
 We can omit these constant factors without prejudice. The SC intensity along $\vq=h\vk{a}^*$ is then 
 just
 \[
 I_{SC}(h)=I_{NC}(h)I_{SL}(h)=U_{N_a-1}^2\lrb{\cos\lrb{\pi h }}U_{M_a-1}^2\lrb{\cos\lrb{\pi h (a_S/a)}}
 \]
 In Fig.~\ref{fig:Laue} we plot both $I_{NC}(h)=U_{N_a-1}^2\lrb{\cos\lrb{\pi h }}$ and 
 $I_{SC}(h)/M_a^2$; the last scaling sets $0<I_{SL}(h)<1$ for convenience. 
 
\clearpage
\section{Superlattice of not identical objects}\label{sec:sizedisSC}

In this section we explore the case when the NCs arranged on the SL are not all equal sized. 
We will call this situation the \emph{size disorder effect} (SDE).

We still assume a paralleloidal shape. 
The dimensional constants $N_a,N_b,N_c$ and the spacing constants $s_a,s_b,s_c$ (see Eq.~\ref{eq:spacy}) may sometimes not be indeed constant and immutable throughout the structure. 
We will suppose instead that all of them may be statistically described by (narrow) distributions over the positive real axis, each having a defined average and variance and all finite superior moments. 
The simplest and most widely used such distributions are lognormals. 
So, for instance, we suppose that $N_a$ has average $\lra{N_a}$ and variance ${\mathcal{V}}_{N_a}$. 
We also introduce for convenience the fractional dispersions 
\[
\eta\equiv \DSF{
\sqrt{{\mathcal{V}}_{N_a}}
}{\lra{N_a}}; \qquad \tau\equiv \DSF{
\sqrt{{\mathcal{V}}_{s_a}}
}{\lra{s_a}}
\]
A lognormal 
probability density describing $N_a$ (represented by the continuous variable $X$) is
\begin{equation}
P_{N_a}(X)=\DSF{1}{X\sqrt{
2\pi\,
\log\lrb{
\lra{N_a}^2\lrb{1+\eta^2}
}
}}\exp\lrs{
-\DSF{1}{2}
\DSF{\lrb{
\log\lrb{X}-\log\lrb{\lra{N_a}}+\lrb{1/2}\log\lrb{1+\eta^2}
}^2
}
{ 
\log\lrb{1+\eta^2}
}
}
\end{equation}
And similarly, for $s_a$, with associated variable $Y$,
\begin{equation}
P_{s_a}(Y)=\DSF{1}{Y\sqrt{
2\pi\,
\log\lrb{
\lra{s_a}^2\lrb{1+\tau^2}
}
}}\exp\lrs{
-\DSF{1}{2}
\DSF{\lrb{
\log\lrb{Y}-\log\lrb{\lra{s_a}}+\lrb{1/2}\log\lrb{1+\tau^2}
}^2
}
{ 
\log\lrb{1+\tau^2}
}
}
\end{equation} 

Similarly for $N_b$, $N_c$ and $s_b$, $s_c$. All the averages are straightly denoted
\[
\lra{N_a},\ \lra{N_b},\ \lra{N_c},\ \lra{s_a},\ \lra{s_b},\ \lra{c_c};
\]
and the variances
\[
{\mathcal{V}}_{N_a},\ {\mathcal{V}}_{N_b},\ {\mathcal{V}}_{N_c},\ {\mathcal{V}}_{s_a},\ {\mathcal{V}}_{s_b},\ {\mathcal{V}}_{s_c}.
\]

\subsection{1-D superlattice}\label{sec:1-DSL}

This is simple set of nanocrystals on a line, hence forming a rod. 
Notation: if a variable $X$ is distributed according to a given probability density $P(X)$ 
whose moments are all finite, 
we denote $\lra{X}$ its average (normalized first moment) and ${\mathcal{V}}_X$ its variance (normalized central moment) under $P(X)$.

Consider first the fully ordered case, where 
\begin{eqnarray}
\lra{N_a}&=&N_a,\qquad {\mathcal{V}}_{N_a}=0;\qquad
\lra{s_a}=s_a,\qquad {\mathcal{V}}_{s_a}=0.\label{case:ord}
\end{eqnarray}

Firstly, for a finite sequence of equispaced points of length $M_a$, 
the multiplicity of the zero distance is $M_a$, while that of any pair of $k$-spaced nodes is
\begin{equation}
{\mu}_{d_k}=2\lrb{M_a-\lrv{k}}
\label{eq:mulD}
\end{equation}

The average distance between two nodes spaced by $k$ superlattice sites will be 
simply 
\[
d_k=k(N_a+s_a)a.
\] 
If now we remove the assumptions in \eref{case:ord}, we have to average over the distribution of \emph{every} 
variable segment. Supposing now every segment is variable.
So,
\begin{eqnarray}
\lra{d_k}&=&a\int\DD{X_1}\int\DD{X_2}\ldots\int\DD{X_k}\ \int\DD{Y_1}\int\DD{Y_2}\ldots\int\DD{Y_k}
\lrb{\mathop{\sum}_{\ell=1}^kX_\ell+\mathop{\sum}_{\ell=1}^kY_\ell}\times\nonumber\\
&\times&P_{N_a}(X_1)P_{N_a}(X_2)\ldots P_{N_a}(X_k)\ P_{s_a}(Y_1)P_{s_a}(Y_2)\ldots P_{s_a}(Y_k)\\
&=&
a\mathop{\sum}_{\ell=1}^k\int\DD{X_\ell}\ X_\ell\ P_{N_a}(X_\ell)+
a\mathop{\sum}_{\ell=1}^k\int\DD{Y_\ell}\ Y_\ell\ P_{S_a}(Y_\ell) \\
&&\text{[because all }P_*\ \text{functions are normalised to 1]}\nonumber\\
&=&k\lrb{ \lra{N_a}+\lra{s_a}} a\qquad \text{by definition}
\label{eq:aveD}
\end{eqnarray}
Similarly, for the variance, repeating similar passages,
we obtain
\begin{equation}
{\mathcal{V}}_{d_k}=\lrv{k}\lrb{ {\mathcal{V}}_{N_a}+{\mathcal{V}}_{s_a}} a^2
\label{eq:varD}
\end{equation}
It is clear that the effect on the interatomic distances of the variability of the size $N_a$ and that of the spacing $s_a$ are indistinguishable. We will consider - unless otherwise specified - a single 
parameter $\eta_a\equiv{N_a}+{s_a}$, so that
\begin{equation}
\lra{d_k}=k\,\lra{\eta_a}a; \qquad {\mathcal{V}}_{d_k}=\lrv{k}\, {\mathcal{V}}_{\eta_a}a^2
\label{eq:aveDvarD_eta}
\end{equation}

\subsection{2-D and 3-D superlattices }

The fully ordered case is described in \sref{sec:perfectSC}. 
We will hereby only consider the orthorhombic case where 
\[
\vk{a}\cdot\vk{b}=\vk{b}\cdot\vk{c}=\vk{c}\cdot\vk{a}=0;\qquad 
\vk{a}_S=\eta_a\vk{a},\qquad
\vk{b}_S=\eta_a\vk{b},\qquad
\vk{c}_S=\eta_a\vk{c}. 
\]
where as before 
\[
\eta_a\equiv{N_a}+{s_a};\qquad \eta_b\equiv{N_b}+{s_b};\qquad \eta_c\equiv{N_c}+{s_c}.
\]

In the ordered case 
\begin{equation}
\lra{\eta_a}=N_a,\quad {\mathcal{V}}_{\eta_a}=0;\qquad
\lra{\eta_b}=N_b,\quad {\mathcal{V}}_{\eta_b}=0;\qquad
\lra{\eta_c}=N_c,\quad {\mathcal{V}}_{\eta_c}=0.\label{case:ord3}
\end{equation}

Consider a SL formed by a parallelogram 
\[
m_a=1,\ldots,M_a;\qquad m_b=1,\ldots,M_b;\qquad m_c=1,\ldots,M_c. 
\]
The vector distance between two SL nodes $\vk{M}=(m_a,m_b,m_c)$ and $\vk{M}'=(m'_a,m'_b,m'_c)$ spaced by 
${\vk{K}}\equiv\lrb{ k_a,k_b,k_c}=(m'_a,m'_b,m'_c)-(m_a,m_b,m_c)$ 
will be 
\[
\vk{d}_{\vk{K}}=k_a\vk{a}_S+k_b\vk{b}_S+k_c\vk{c}_S=
\lrb{ k_a\eta_aa,k_b\eta_bb,k_c\eta_cc}
\]
And 
it is immediate to generalise \eref{eq:mulD} for the multiplicity of $\vk{d}_{\vk{K}}$
as 
\begin{equation}
\mu_{\vk{d}_{\vk{K}}}=\lrb{M_a-\lrv{k_a}}\lrb{M_b-\lrv{k_b}}\lrb{M_c-\lrv{k_c}}.
\label{eq:mulD3}
\end{equation}

The total distance between two point scatterers belonging each to one of the two NC centered at the $\vk{M}$ and $\vk{M}'$ SL nodes must also take into account the difference between respective position vectors $\vk{n}=\lrb{n_a,n_b,n_c}$ and $\vk{n}'=\lrb{n'_a,n'_b,n'_c}$ in the 
generic NC lattice $\vk{g}\equiv\vk{n}-\vk{n}'$. It results
\begin{equation}
\vk{d}_{\vk{K},\vk{g}}=\vk{d}_{\vk{K}}+\vk{d}_{\vk{g}}=\vk{d}_{\vk{K}}+g_a\vk{a}+g_b\vk{b}+g_c\vk{c}=
\lrb{ k_a\eta_aa,k_b\eta_bb,k_c\eta_cc}+\lrb{ g_aa,g_bb,g_cc}
\label{eq:fullD}
\end{equation}

If we consider instead the $\eta$ parameters to follow a probability density with all finite moments, 
we can repeat the calculations in \sref{sec:1-DSL} component by component. 
We have to add an assumption - that the joint distribution is the product of the single variable distributions, or 
\[
P_{\eta_a,\eta_b,\eta_c}(X_a,X_b,X_c)=P_{\eta_a}(X_a)P_{\eta_b}(X_b)P_{\eta_c}(X_c)
\]
This will cause the covariance to be diagonal. Removing this assumption is straightforward, 
but it leads to far more complex bookkeeping. 

We have then the vector average
\begin{equation}\label{eq:vecAv}
\lra{\vk{d}_{\vk{K}}}=\lrb{ k_a\lra{\eta_a}a,k_b\lra{\eta_b}b,k_c\lra{\eta_c}c}
\end{equation}
The NC-related distance vector $\vk{d}_{\vk{g}}$ in \eref{eq:fullD} is constant, therefore it adds to the average and does not contribute to the variance. The averages result to
\begin{equation}\label{eq:vecAv}
\lra{\vk{d}_{\vk{K},\vk{g}}}=\lrb{ \lrb{ k_a\lra{\eta_a}+g_a}a,\lrb{k_b\lra{\eta_b}+g_b}b,\lrb{k_c\lra{\eta_c}+g_c}c}
\end{equation}
And we have a diagonal covariance
matrix ${\bm{\mathcal{V}}}_{\vk{d}_{\vk{K},\vk{g}}}$, that is actually independent on $\vk{g}$:
\begin{equation}\label{eq:vecCov}
{\bm{\mathcal{V}}}_{\vk{d}_{\vk{K},\vk{g}}}=\lrb{ 
\begin{array}{ccc}
\lrv{k_a}\mathcal{V}_{\eta_a}  & 0  & 0  \\
 0 & \lrv{k_b}\mathcal{V}_{\eta_b}  & 0  \\
 0 & 0 & \lrv{k_c}\mathcal{V}_{\eta_c}    
\end{array}
}
\end{equation}

We cannot be too specific on the form of the 3-D distribution of $\vk{d}_{\vk{K}}$; however, 
it is not wrong to assume it being a 3-D Gaussian with specified averages and covariance matrix. 
Then we would have
\begin{equation}
P\lrb{\vk{d}_{\vk{K},\vk{g}}} = \DSF{1}{(2\pi)^{3/2}\sqrt{\det{{\bm{\mathcal{V}}}_{\vk{d}_{\vk{K},\vk{g}}}}}}
\exp\lrs{-\DSF{1}{2} \lrb{{\vk{d}_{\vk{K},\vk{g}}}-\lra{\vk{d}_{\vk{K},\vk{g}}}}
\cdot {\bm{\mathcal{V}}}_{\vk{d}_{\vk{K},\vk{g}}} \lrb{{\vk{d}_{\vk{K},\vk{g}}}-\lra{\vk{d}_{\vk{K},\vk{g}}}} }
\label{eq:d3d_dist}
\end{equation}

\subsection{Powder diffraction signal: powder average} 

Powder average is the average of the diffraction pattern over all possible orientations in space with an uniform distribution. The result will be a function only of $q=\lrv{\vq}$, and it will depend only on the lengths of the interatomic distances. 

For a system of $N$ atoms (simplified as point scatterers) with coordinates $\vkr_j$, $j=1,\ldots,N$, 
each with scattering length $b_j$, 
the powder averaged intensity (differential cross section) can be written 
by means of the the Debye scattering equation \cite{Debye1915} (hereafter DSE) 
as
\begin{equation}
I(q)=\mathop{\sum}_{j,k=1}^{N} b_j b_k \sinc{\lrb{2\pi q\lrv{\vkr_j-\vkr_k}}}=
\mathop{\sum}_{j=1}^Nb_j^2 + \mathop{\sum}_{j \neq k=1}^{N} b_j b_k \sinc{\lrb{2\pi qd_{jk}}}
\label{eq:DSEmu1}
\end{equation}
with $\sinc(x)=sin(x)/x$ is the sine cardinal function and 
where we set $d_{jk}\equiv\lrv{\vkr_j-\vkr_k}$.

For periodically ordered systems, where many of the distances $d_{jk}$ 
will be the same, and also the scattering lengths pair is the same. Then we can group terms in the left sum, leaving $M_d$ distinct $d$-values, each with a multiplicity $\mu$. Then we can write
\begin{equation}
I(q)=\
\mathop{\sum}_{j=1}^{N} b_j^2 + \mathop{\sum}_{\ell=1}^{M_d}
b^2_\ell \mu_\ell\sinc{\lrb{2\pi qd_{\ell}}}
\label{eq:DSEmu}
\end{equation}
If the system is slightly disordered, the $\ell$-indexed groups of $\mu_\ell$ distances 
might become slightly spread in value. If the spread is relatively small, we can refrain from breaking the 
$\ell$-groups and instead evaluate the group average $\lra{d_\ell}$ and its variance $\mathcal{V}_{d_\ell}$. 
Then an effective way of modifying \eref{eq:DSEmu} has been derived \cite{ACNM_RuCO}, with excellent approximation (see also \cite{HosemannBagchi_1962,Welberry_2004}; this case corresponds to a paracrystalline type of disorder with no 
cross-interactions and with positive full correlation (value 1) along each axis. Correlation values below 1 would mean that the NC 
and the spacer would deform elastically to try to partially accommodate differences in size. This is a possible generalisation of this work, but we will not pursue it here as we deem it likely to be of minor importance.) 
The modified DSE reads
\begin{equation}
I(q)=\
\mathop{\sum}_{j=1}^{N} b_j^2 + \mathop{\sum}_{\ell=1}^{M_d}
b^2_\ell \mu_\ell\sinc{\lrb{2\pi q\lra{d_{\ell}}}}\exp\lrb{-2\pi^2q^2\mathcal{V}_{d_\ell}}
\label{eq:DSEmuvar}
\end{equation}
The exponential factor is the Fourier transform of a Gaussian with variance $\mathcal{V}_{d_\ell}$. 

We recall briefly that the DSE is lust the spherical average (over all possible orientations, with uniform distribution) 
of the 3-D scattering equation
\begin{equation}
I(\vq)=\
\mathop{\sum}_{j=1}^Nb_j^2 + \mathop{\sum}_{\ell=1}^{M'_d}
b^2_\ell \mu'_\ell\cos\lrb{2\pi \vq\cdot\lra{\vk{d}_{\ell}}}\exp\lrb{-2\pi^2\vq\cdot{\bm{\mathcal{V}}}_{\vk{d}_\ell}\vq}
\label{eq:3dsceq}
\end{equation}
where the multiplicities may differ (coincidences in 3-D space are more rare). This equation is usually obtained as the square modulus of the direct Fourier transform of the scattering density. 

Suppose now that we have a distribution for 3-D vector distance with a vector average and a covariance matrix (as in 
\eref{eq:d3d_dist}). Knowing $ \lra{\vk{d}_{\vk{K}}}$ (\eref{eq:vecAv}) and the covariance 
${\bm{\mathcal{V}}}_{\vk{d}_{\vk{K}}}$ (\eref{eq:vecCov}), and being
\[
{d}_{\vk{K},\vk{g}}=\sqrt{\lra{\vk{d}_{\vk{K},\vk{g}}}\cdot\lra{\vk{d}_{\vk{K},\vk{g}}}}=\lrb{
(k_a\lra{\eta_a}+g_a)^2a^2+
(k_b\lra{\eta_b}+g_b)^2b^2+
(k_c\lra{\eta_c}+g_c)^2c^2}^{1/2}
\]
where the leftmost expression comes from \eref{eq:vecAv},
we must evaluate the latter's average and variance over the 3-D distribution \eref{eq:d3d_dist}. 
\begin{eqnarray}
\lra{{d}_{\vk{K},\vk{g}}}&=&\int_{\mathbb{R}^3}\DDD{3}{\vk{d}_{\vk{K},\vk{g}} }
P\lrb{\vk{d}_{\vk{K},\vk{g}}} \ d_{\vk{K},\vk{g}};\\
\mathcal{V}_{{d}_{\vk{K},\vk{g}}}
&=&\int_{\mathbb{R}^3}\DDD{3}{\vk{d}_{\vk{K},\vk{g}} }
P\lrb{\vk{d}_{\vk{K},\vk{g}}} \ \lrb{d_{\vk{K},\vk{g}}-\lra{{d}_{\vk{K},\vk{g}}}}^2
\end{eqnarray}
The integrals are not analytic but a series expansion of the integrands 
to the second order around the averages by component of 
${d}_{\vk{K},\vk{g}}$ yields
\begin{eqnarray}
\lra{{d}_{\vk{K},\vk{g}}} &=& 
{d}_{\vk{K},\vk{g}} + \DSF{{d}_{\vk{K},\vk{g}}^2 A_{\vk{K},\vk{g}}-B_{\vk{K},\vk{g}} }{2 {d}_{\vk{K},\vk{g}}^3};\label{eq:dave}\\
\mathcal{V}_{{d}_{\vk{K},\vk{g}}}
&=&
\DSF{B_{\vk{K},\vk{g}}}{{d}_{\vk{K},\vk{g}}^2} - 
\DSF{
\lrb{
{d}_{\vk{K},\vk{g}}^2 A_{\vk{K},\vk{g}}-B_{\vk{K},\vk{g}}}^2 }{4 {d}_{\vk{K},\vk{g}}^6}
\label{eq:dvar}
\end{eqnarray}
where 
\begin{eqnarray}
A_{\vk{K},\vk{g}}&\equiv&
\text{Tr}\lrb{{\bm{\mathcal{V}}}_{\vk{d}_{\vk{K},\vk{g}}}}=\lrv{k_a}\mathcal{V}_{\eta_a}+\lrv{k_b}\mathcal{V}_{\eta_b}+\lrv{k_c}
\mathcal{V}_{\eta_c};\\
B_{\vk{K},\vk{g}}&\equiv&{\vk{d}_{\vk{K},\vk{g}}}\cdot
{\bm{\mathcal{V}}}_{\vk{d}_{\vk{K},\vk{g}}}
{\vk{d}_{\vk{K},\vk{g}}}=
\lrv{k_a}^3\lra{\eta_a}^2a^2\mathcal{V}_{\eta_a}+
\lrv{k_b}^3\lra{\eta_b}^2b^2\mathcal{V}_{\eta_b}+
\lrv{k_c}^3\lra{\eta_c}^2c^2\mathcal{V}_{\eta_c}
\end{eqnarray}
We only then have to plug the $\lra{{d}_{\vk{K},\vk{g}}}$ and $\mathcal{V}_{{d}_{\vk{K},\vk{g}}}$ from \eeref{eq:dave}{eq:dvar} in \eref{eq:DSEmuvar} in place of 
$\lra{d_\ell}$ and of $\mathcal{V}_{d_\ell}$, respectively. The multiplicity $\mu_\ell$ is given in \eref{eq:mulD3}.

\section{Superlattice of misaligned objects}

We explore also - partly - the case when the NCs arranged on the SL are all equal sized 
(no SDE) but not perfectly aligned with each other. This we name the \emph{alignment disorder effect} (ADE).

We develop this case very briefly because of the extensive theoretical analysis involved, that suggests to dedicate a specific manuscript to it. However, we want to give at least a feeling of the effect on diffrection of alignmet disorder.

Take two SL sites separated by $\vk{K}=\lrb{K_a,K_b,K_c}$ nodes, the actual displacement vector being 
$K_a\vk{a}_S + K_b \vk{b}_S + K_c\vk{c}_S$. 
One NC at one end of $\vk{K}$ is held fixed, an identical one at the other end is subjected to a general rotation. 
A general rotation in 3-D space can be described as three subsequent rotations along three non-coplanar directions; 
for convenience we choose the directions of $\vk{a}_S,\vk{b}_S,\vk{c}_S$ as axes, in the order. 
The rotations are quantified by three angles $\phi_{\vk{a}_S}$, $\phi_{\vk{b}_S}$, $\phi_{\vk{c}_S}$, respectively. 

As for the size disorder case, we imagine an equivalent mechanism where nearest-neighbour only interactions are involved. 
As such, every NC has a small rotational degree of freedom with respect to its nearest neighbours. 
The variances of the rotation angles then increases linearly with the number of steps in each SL direction between two SL sites.
It is reasonable that each SL direction influences differently the rotation angle around itself than the other SL directions.
Then we have a simple matrix equation for evaluating the angular variances,
\begin{equation}
\label{eq:rotdef}
\left(
\begin{array}{c}
  \mathcal{V}_{\phi_{\vk{a}_S} }  \\
   \mathcal{V}_{\phi_{\vk{b}_S} } \\
     \mathcal{V}_{\phi_{\vk{c}_S} }
\end{array}
\right)
=
\left(
\begin{array}{ccc}
 \xi & \chi  & \chi  \\
 \chi & \xi  & \chi  \\
 \chi & \chi  & \xi  
\end{array}
\right)
\left(
\begin{array}{c}
  K_a   \\
  K_b   \\
  K_c  
\end{array}
\right)
\end{equation}
We require also to have no net rotation, or equivalently zero angle averages 
$\lra{\phi_{\vk{a}_S} }=\lra{\phi_{\vk{b}_S} }=\lra{\phi_{\vk{c}_S} }=0$. 

The two NC spaced by $\vk{K}$ have each a diffraction amplitude $F_{NC}(\vq)$
described by \eref{eq:LaueANC}. 
For the one NC that is rotated, also $F_{NC}(\vq)$ will be rotated; we 
indicate it simply as $F'_{NC}(\vq)$. The total diffraction amplitude is then
\begin{equation}
F_{NC}(\vq)+\exp\lrb{
2\pi\IMA \vq\cdot
\lrb{
K_a\vk{a}_S
+K_b\vk{b}_S
+K_c\vk{c}_S
}
}
F'_{NC}(\vq)
\end{equation}
The intensity will be its square modulus
\begin{equation}
F^2_{NC}(\vq)+F'^2_{NC}(\vq)+2F_{NC}(\vq)F'_{NC}(\vq)\cos\lrb{2\pi\vq\cdot\lrb{
K_a\vk{a}_S
+K_b\vk{b}_S
+K_c\vk{c}_S
}}
\end{equation}
The term containing the product $F_{NC}(\vq)F'_{NC}(\vq)$ will be greatly reduced because the rotation 
will cause peaks of $F'_{NC}(\vq)$ to rotate out of the corresponding peaks of $F_{NC}(\vq)$ (except the origin peak, 
that is only relevant for SAXS, of course). 
The most dramatic effect will be 
when even the lowest lying peaks are totally decoupled. 
Supposing $a=b=c$ (cubic NC cell) and $N_a=N_b=N_C$ (cubic NC), as the footprint of a peak in each direction 
extends from $-(aN_a)^{-1}$ to $(aN_a)^{-1}$, the rotation angle necessary to maximally suppress the first (100) peak located at $q=1/a$
will be $\Phi\approx\arctan\lrb{2/N_a}$.
This gives us a criterion for understanding when a rotation is small or disruptively large. 
Cumulative effects will be explored elsewhere [fig arriving].

\section{Example calculations}\label{sec:ex}

Here we want to show some numerical calculations of SC diffraction patterns with size disorder effect (SDE). 
We will start with a system that produces truly 1-D scattering (a set of parallel planes does that). Then we will have 
SCs with small NCs and different degrees of disorder and also different SL dimensionality (rods, planes, and true bulk SC). 

\subsection{1-D chain of parallel planes with 1-D scattering} 

This case represenrts the practical case f a set of parallel planes whose diffraction is measured in $\vq$-space along the direction orthogonal to the planes.

\begin{figure}[!htb]
\centering
 \includegraphics[width=10.5cm]{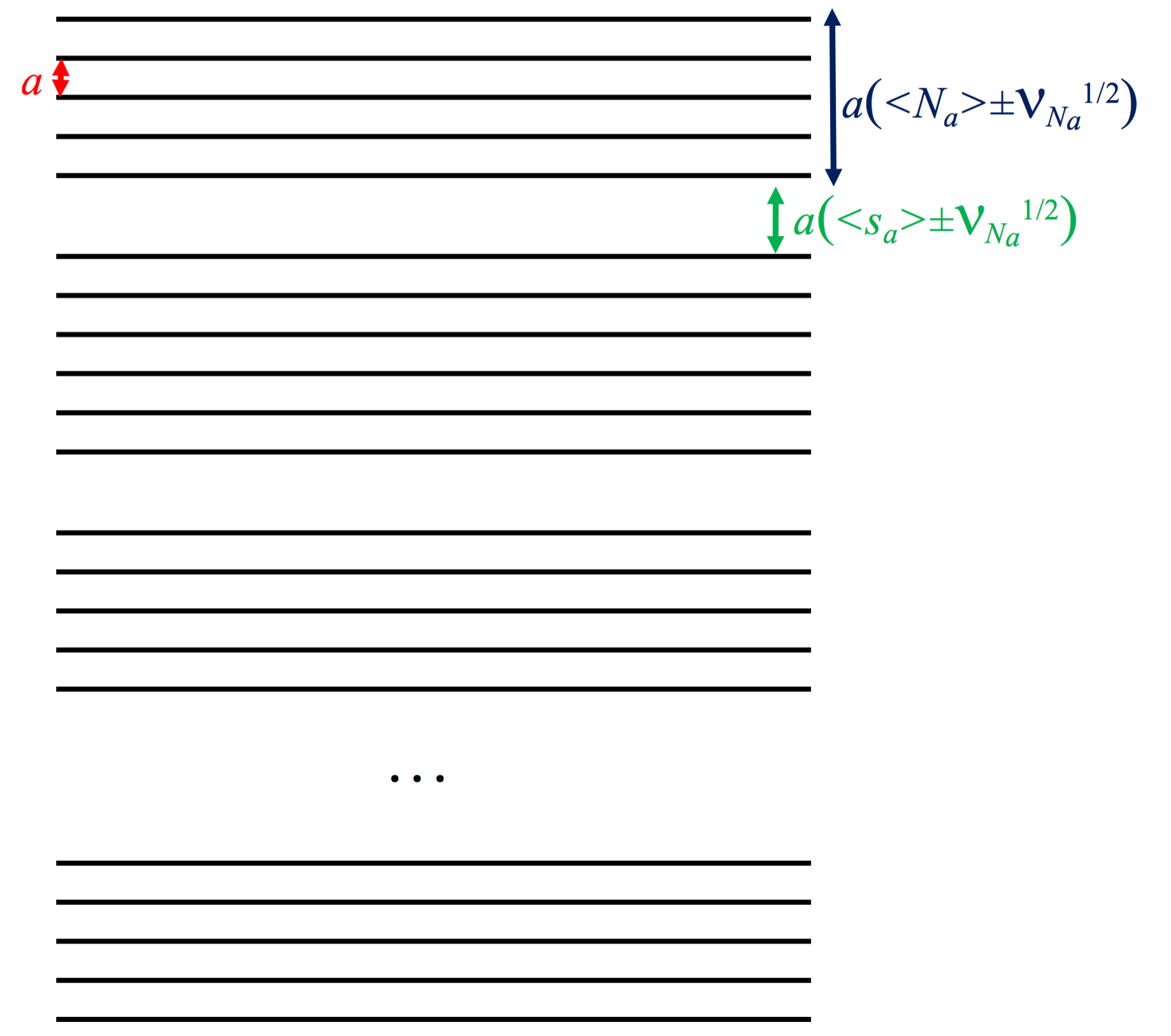}
\caption{This is a representation of a system with 1-D SDE. It consists of a stacking of parallel planes, sub-ordered in bunches of 
different height. Planes (orthogonal to the figure) are represented by their traces; the vertical direction is the stacking direction, normal to the planes. The spacing between planes in a bunch is $a$, inter-bunch spacing 
$a\lrb{\lra{s_a}\pm\mathcal{V}^{1/2}_{s_a}}$; 
each bunch consists of a number of planes that in average is $\lra{N_a}$ with a dispersion $\mathcal{V}^{1/2}_{N_a}$. }
\label{fig:geom}
\end{figure}

It is noteworthy that the dimensions orthogonal to the stacking direction (that is normal to the planes) can be ignored.

The scattering equation of this system reads
\[
I(q)=\lrv{
\mathop{\sum}_{m=1}^{M_a}\exp{\lrb{2\pi \IMA q z_m}}
}^2
=M_a+\mathop{\sum}_{m\ne m'=1}^{N_{stack}}\cos{\lrb{2\pi q {\lrv{z_m-z_{m'}}}}}
\]
where $z_m$ are the coordinates along the stacking axis $z$, $q$ is the scattering vector along the same axis, 
and $M_a$ the total number of planes. It is similar to the DSE \eref{eq:DSEmu1}
where the $\Sinc$ function is replaced by a more mundane cosine. 

Each bunch of planes is equispaced, so we can write, for an isolated bunch of height $N_a$ and spacing $a$, 
(the latter we suppose to be the same for all bunches, the former we let be variable)
\[
I_{bunch}(q,N_a)=\mathop{\sum}_{k =  -N_a+1}^{N_a-1}\lrb{N_a-|k|}
\cos\lrb{2\pi qka}
\]
If we average over a bunch size distribution on a finite discrete range
\[
P(N_a), \qquad N_a=1,\ldots,\widehat{N}_a \qquad\left|\qquad  \mathop{\sum}_{N_a=1}^{\widehat{N}_a}
P(N_a) = 1\right.
\]
with
\[
\lra{N_a}=\mathop{\sum}_{N_a=1}^{\widehat{N}_a}
N_aP(N_a); \qquad \mathcal{V}_{N_a}=\mathop{\sum}_{N_a=1}^{\widehat{N}_a}
\lrb{N_a-\lra{N_a}}^2P(N_a),
\]
then we can write, for the average bunch,
\begin{equation}
I_{bunch}(q)=
\mathop{\sum}_{k =  -\widehat{N}+1}^{\widehat{N}-1}
\lrs{\mathop{\sum}_{N_a=1}^{\widehat{N}_a}P(N_a)
\max\lrb{0,N_a-|k|}}
\cos\lrb{2\pi qka}
\equiv\mathop{\sum}_{k =  -\widehat{N}+1}^{\widehat{N}-1}\mu_k\cos\lrb{2\pi qka}
\end{equation}

The average SL 
is the periodic average of the arrangement of bunches, with an average spacing 
$a\eta_a=a\lrb{N_a+s_a}$.
The scattering from a SL of $M_a$ plane bunches then results to
\begin{eqnarray}
I(q)&=&\mathop{\sum}_{m=-M_a+1}^{M_a-1}\ 
\mathop{\sum}_{k =  -\widehat{N}_a+1}^{\widehat{N}_a-1}\mu_k\,
\lrb{M_a-|m|}
\ \cos\lrb{2\pi qa\lrv{
\lrb{ 
k+ m\lrb{ \lra{N_a}+\lra{s_a} } } 
}
}  
\times\nonumber\\&\times& \exp\lrs{
-2m\pi^2q^2a^2\lrb{ {\mathcal{V}}_{N_a}+{\mathcal{V}}_{s_a}} 
}\nonumber
\end{eqnarray}

Example 1-D patterns. We consider bunches of equispaced planes (representing the NCs) stacked with 
dead space on top of each other, see \fref{fig:geom}.
The distribution is nonzero only at two values:
\[
\begin{array}{ccl}
1<N_a<5 & \rightarrow & P(N_a)=0\\
N_a=5 & \rightarrow & P(5)=0.713384\\
N_a=6 & \rightarrow & P(6)=0.286616\\
N_a>6 & \rightarrow & P(N_a)=0
\end{array}
\]
resulting in
\[
\begin{array}{ccc}
\lra{N_a}=5.286616\\
{\mathcal{V}}_{N_a}=0.204467\\
\DSF{\sqrt{{\mathcal{V}}_{N_a}}}{\lra{N_a}}=0.0855331 \quad(\approx 8.5\%)
\end{array}
\]
As we see, this case results in a "narrow" distribution with 
8.5\%\ relative dispersion.

\begin{figure}[!htb]
\centering
\includegraphics[width=10.5cm]{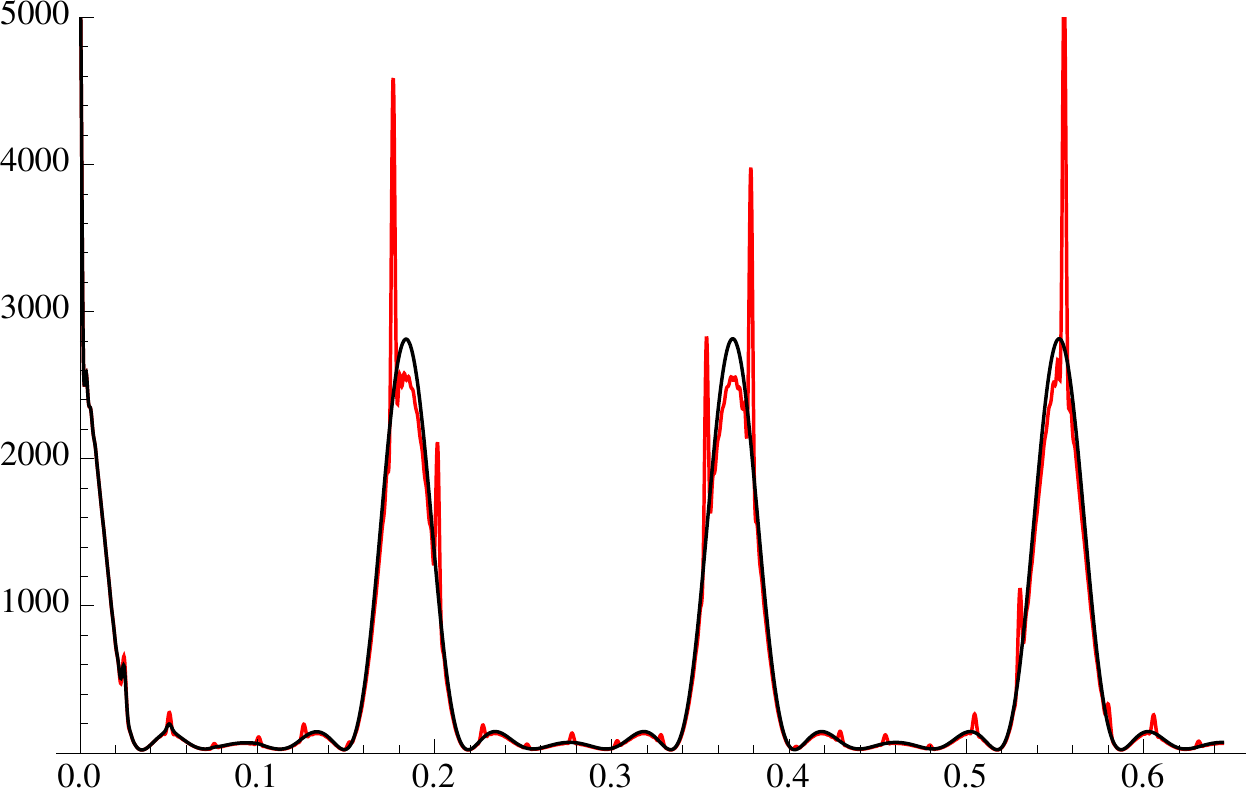}
\caption{Plane bunch sequence 1-D diffraction pattern: Red - calculation with zero spacing dispersion; Black - including 8.5\%\ spacing dispersion as from the text. The 8.5\%\ dispersion destroys the sharp small peaks 
(SL interference) except in the small-angle region.}
\label{fig:culo}
\end{figure}

We also set 
$a=5.431\ \text{\AA}$. 
We set the interbunch spacing to $0.38 \lra{N_a} a$. This we suppose to have zero variance. 
We take $M_a=10$. 
In \fref{fig:culo} we see calculated diffraction patterns - switching on and off the 8.5\%\ spacing dispersion

\begin{figure}[!htb]
\centering
 \includegraphics[width=10.5cm]{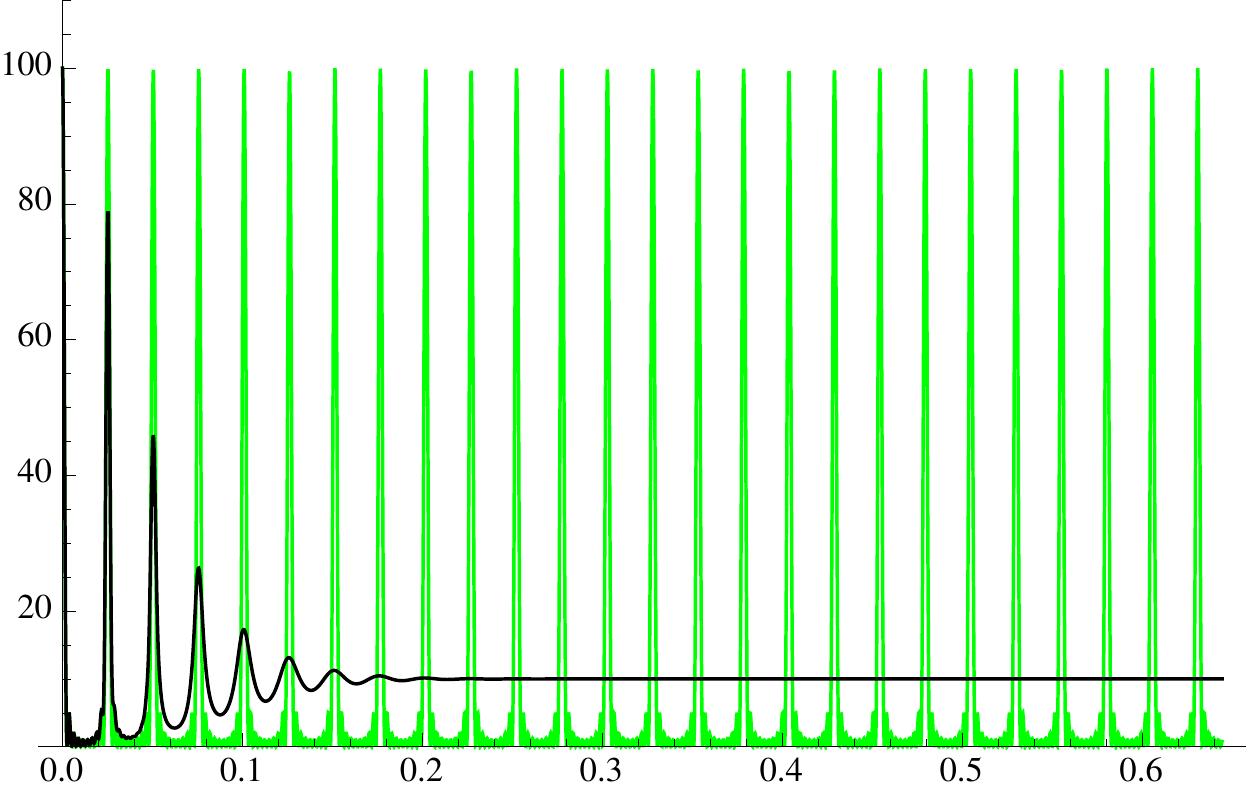}
\caption{Basic 1-D diffraction pattern of the same sequence, just substituting the plane bunches with single plane scatterers - to see the naked decoherence effect of the spacing dispersion. Green - zero dispersion, black - 8.5\%\ dispersion}
\label{fig:ossi}
\end{figure}

We also calculated the diffraction pattern in the case where every plane bunch (or NC) is substituted by a single scattering plane. 
This shows directly (\fref{fig:ossi}) the SL scattering and the interference (or lack thereof) when the spacing is subjected to the same 8.5\%\ dispersion. 

\subsection{1-D, 2-D and 3-D SCs}

\begin{figure}[!htb]
\centering
 \includegraphics[width=10.5cm]{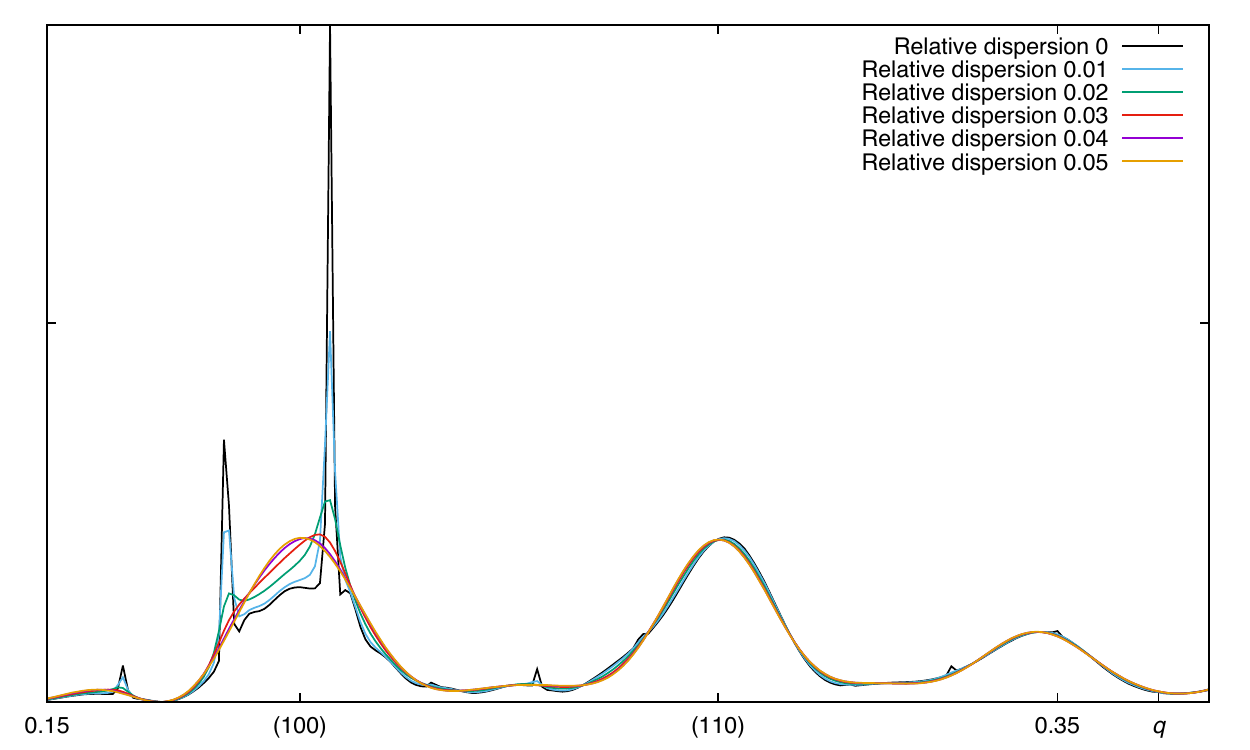}
\caption{XRPD scattering from a rod (1-D SC) with various level of relative SL spacing dispersion.}
\label{fig:rod}
\end{figure}

\begin{figure}[!htb]
\centering
 \includegraphics[width=10.5cm]{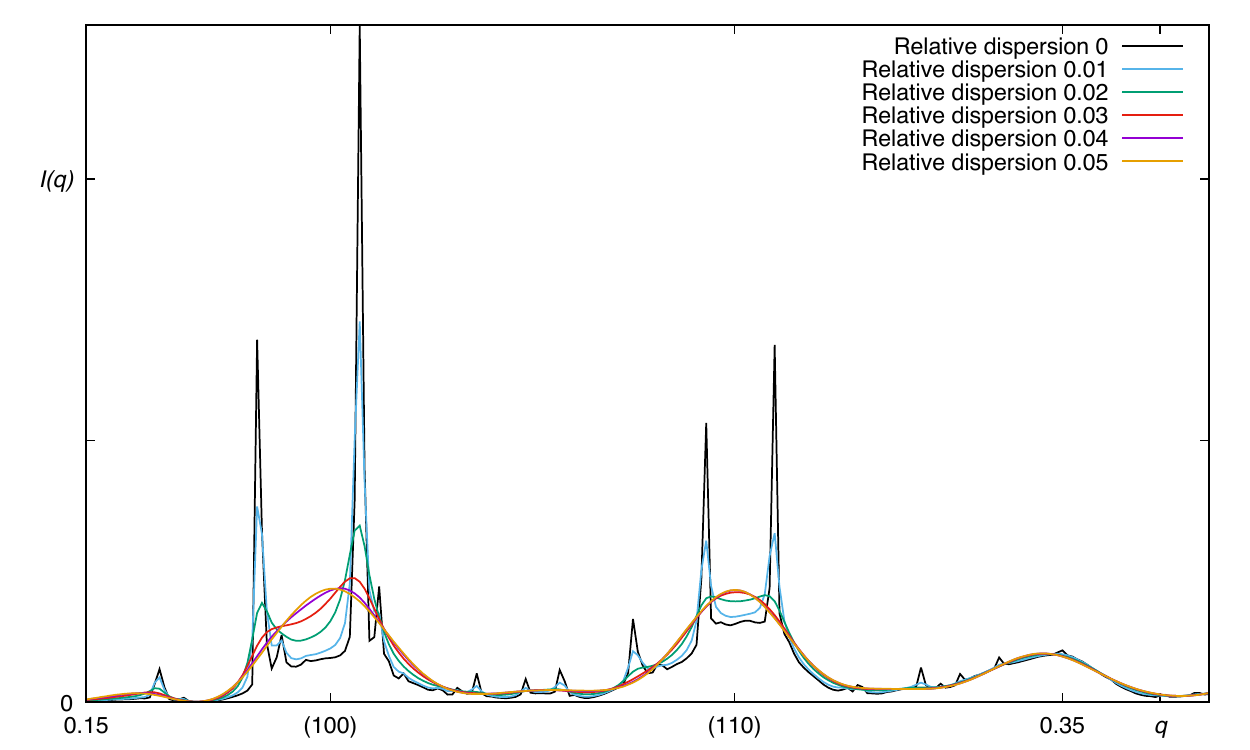}
\caption{XRPD scattering from a plate (2-D SC) with various level of relative SL spacing dispersion.}
\label{fig:plate}
\end{figure}

\begin{figure}[!htb]
\centering
 \includegraphics[width=10.5cm]{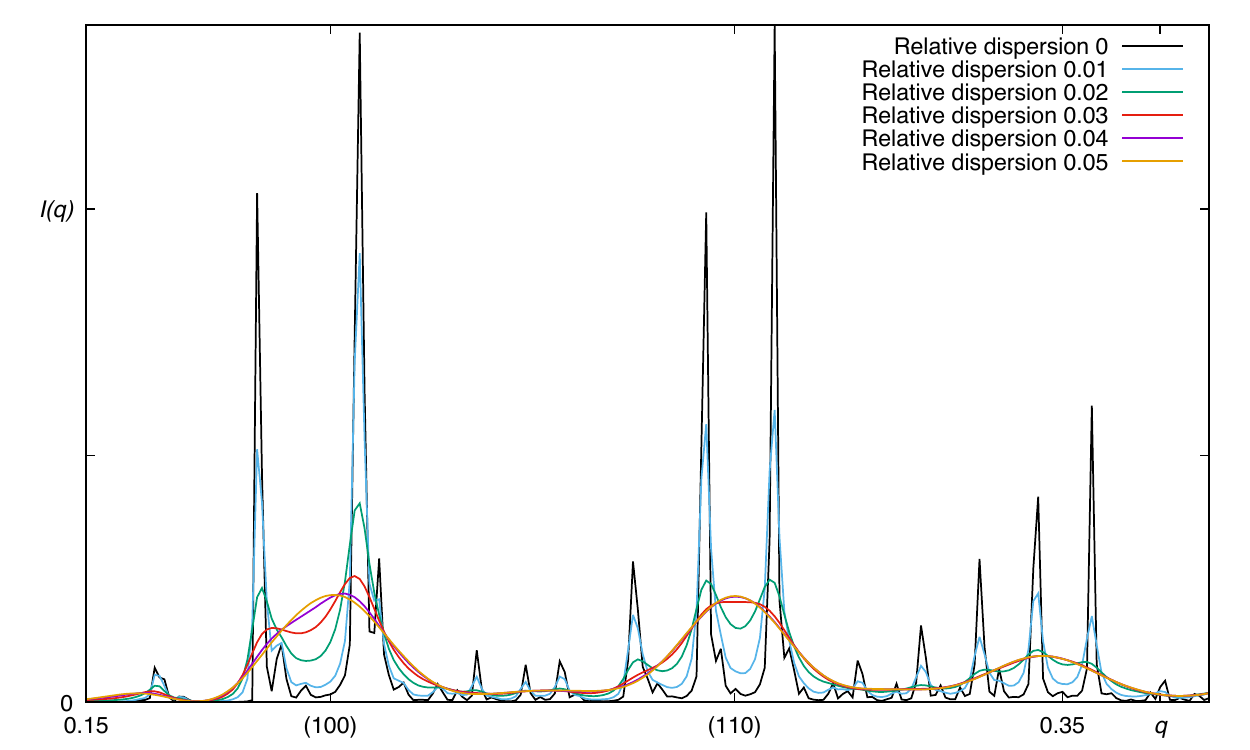}
\caption{XRPD scattering from a cube (3-D SC) with various level of relative SL spacing dispersion.}
\label{fig:cube}
\end{figure}

We constructed cubic NCs (lattice parameter $a=5$~\AA) with $7\times 7\times 7$ unit cells (each cell containing just one point scatterer of unitary length) and arranged them on large cubic SLs with 
superlattice parameter $a_S=9.73 a$. 
The SL dimensions were $20\times 1 \times 1$ unit cells (a rod-like or 1-D SC), 
 $20\times 20 \times 1$ unit cells (a plate-like or 2-D SC), 
  $20\times 20 \times 20$ unit cells (a cube-like or 3-D SC).
 In all cases, the 1-D powder diffraction trace was evaluated in a wide range with 
 different settings of the spacing dispersion (0\%, 1\%, 2\%, 3\%, 4\%, 5\%). 
 Interesting details of calculated traces are shown in \fref{fig:rod} (for the rod), \fref{fig:plate} (for the plate) 
 and \fref{fig:cube} for the cube. Note a general reinforcement of the SL interference scattering (sharp features), 
 and also note how in general a small fractional dispersion (below 5\%) is always able to destroy the SL interference
 on all NC Bragg peaks.

\newpage
 
\subsection{Atomic simulations and 3-D scattering}
\begin{figure}[!htb]
\centering
 \includegraphics[width=0.9\textwidth]{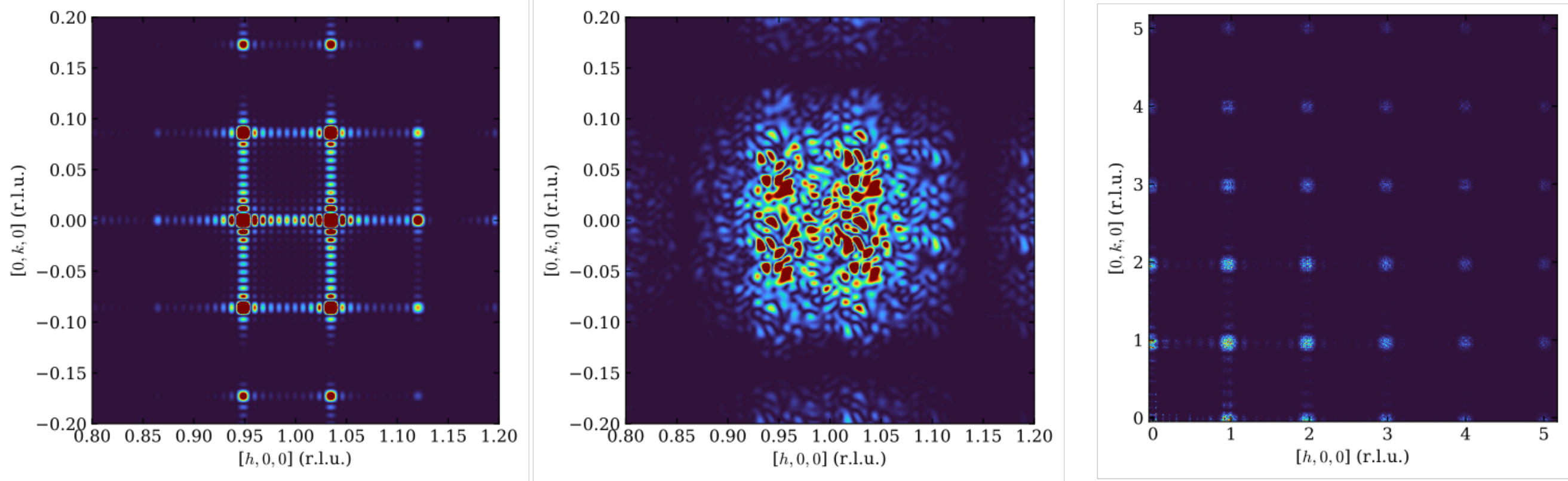}
\caption{2D Sections of the reciprocal lattice of the ideal (left) and size-disordered SL (middle and right). Left and middle: detail of the 100 peak in r.l.u. of the NC lattice. Right: section of the ($hk0$) plane of the size-disordered SL, showing a strong decay with $\vq$. All plots are in the same scale}
\label{fig:scatt2D}
\end{figure}

We computed the diffracted intensity (as from\eref{eq:3dsceq}) in the ($hk0$) reciprocal lattice plane of an 11x11x1 SL of Ni NCs of 7x7x7 unit cells. In \fref{fig:scatt2D} the diffraction patterns of both ideal (\sref{sec:perfectSC}) and SDE-affected SC (\sref{sec:sizedisSC}) are shown. In the first case we computed a superlattice model crystal composed of identical NC regularly spaced, in the second a fraction of the NC's in the ideal SL was replaced with larger NC crystals and the NC's centre-to-centre distance adjusted in order to preserve the NC's spacing.

\section{Discussion}

We have explored the peculiar disorder effects of NC-based SCs, 
that constitute a growing trend because the SL order introduces or modifies physical properties in 
ways that are interesting for applications. To explore the underlying mechanism, a great importance is 
attached to the fine structural features of the superorder. As a rule of thumb, 
structural effects that modify the X-ray diffraction are also modifying the electronic properties through 
the band structure. This makes it interesting to explore the quality requirements - in terms of 
NC size and shape uniformity, and also in terms of co-alignment of the NCs until their periodic arrangement on a SL 
truly forms a SC. To this aim, we have investigated the diffraction footprint of 
SC whose constituting NCs have a small size dispersion that must affect the quality of the periodic SL order.
It turns out that a small size dispersion (4-5\%) is already able to severely affect (up to canceling) the SL coherence, 
whilst NC misalignment is also very effective at this task but its destructive effect is higher for larger NC sizes. 
Therefore, in order to achieve SL interference effects - if they are connected to desirable changes in the electronic properties - 
a great care must be taken to ensure a very sharp size distribution (with relative 
dispersion at the \%\ order) and a great uniformity and regularity of shape (in the reasonable hypothesis that 
large flat NC facets would hider misalignments).

\end{document}